\documentclass[12pt]{iopart}
\usepackage{cite}
\usepackage{amssymb}
\usepackage{amsmath}
\usepackage{bm}
\usepackage{epsfig}
\usepackage{iopams}

\usepackage[colorlinks=true, allcolors=black]{hyperref}
\usepackage{xcolor}

\begin{document}

\title{Generating functions and Bloch representation for quantum metric, Berry curvature, and quantum Fisher tensor}




\author{Felipe P. Abreu and Wei Chen}

\address{Department of Physics, PUC-Rio, Rio de Janeiro 22451-900, Brazil}

\ead{fabreu@aluno.puc-rio.com, wchen@puc-rio.br}

\begin{abstract}

The Uhlmann relative amplitude between two density matrices is shown to be a generating function, through which the quantum Fisher tensor that contains both the quantum Fisher information matrix and the mean Uhlmann curvature can be obtained via differentiation over system parameters. In the pure state limit, our generating function recovers that of the quantum geometric tensor proposed by Het\'{e}nyi and L\'{e}vay, and also clarifies the fidelity between two quantum states as a generating function for quantum metric, and the Pancharatnam phase as a generating function for Berry connection and Berry curvature, as elaborated by the Su-Schrieffer-Heeger model and Chern insulator. A generic expression for the quantum Fisher tensor in terms of the Bloch representation of density matrices is derived, which facilitates the calculation of the tensor, mean Uhlmann curvature, and geometric properties derived from the quantum Fisher information matrix. Canonical ensembles of spins are adopted as an example, which reveals a constant Ricci scalar, a vacuum Einstein equation, and a cosmological constant on the 3D Euclidean manifold of the magnetic field.  

\end{abstract}

\section{Introduction}


A key ingredient in the field of quantum metrology is the Quantum Fisher Information Matrix (QFIM)\cite{Helstrom67,Hovelo11,Braunstein94,Braunstein96,Paris09,Liu20}. This quantity is introduced in a way analogous to the classical FIM, defined as the variance of the score of a probability density function\cite{Fisher25}, which may be viewed as a metric that characterizes the curved parameter space as a Euclidean manifold, giving rise to the notion of information geometry. The generalization of this concept to quantum systems utilizes the density matrix and the associated symmetric logarithmic derivative (SLD) that serves as the quantum analog of the score, leading to the QFIM that manifests in all different branches of quantum metrology, such as quantum sensing\cite{Degen17}, quantum speed limit\cite{Deffner17}, quantum thermodynamics\cite{Vinjanampathy16}, etc. Most notably, just like the Cram\'{e}r-Rao bound derived from classical FIM that limits the accuracy of any classical estimators to system parameters\cite{Rao45,Cramer46}, the quantum Cram\'{e}r-Rao bound derived from QFIM also limit the precision of quantum estimators, indicating the importance and ubiquity of QFIM\cite{Helstrom67,Hovelo11,Braunstein94,Braunstein96,Paris09,Liu20}. Further limiting the discussion to pure states, the QFIM reduces to the quantum metric\cite{Provost80}, also known as the fidelity susceptibility\cite{You07,Zanardi07}, which characterizes the distance between two quantum states that are infinitely close in parameter space. A wide variety of physical properties of solids, such as dielectric and optical properties of semiconductors\cite{Ozawa18,Ahn22,Komissarov24,Chen25_optical_marker}, are recently shown to be originating from the quantum metric, which has stimulated a great deal of interest in this quantity.

Another ingredient that is also of tremendous importance in quantum metrology is the mean Uhlmann curvature (MUC), which characterizes the Uhlmann phase accumulated by bringing a density matrix along an infinitesimal loop in the parameter space\cite{Uhlmann86,Uhlmann91,Carollo18}. This curvature is the mixed state analog of the celebrated Berry curvature, which is the mechanism behind numerous phenomena\cite{Berry84}, such as the Aharonov$-$Bohm effect\cite{Aharonov59} and quantized Hall conductance\cite{Thouless82}, just to name a few. Furthermore, MUC can also be expressed in terms of the SLD, indicating its close relation to the QFIM. In fact, just like the quantum metric and Berry curvature are the real and imaginary parts of a more generalized geometric quantity called quantum geometric tensor\cite{Provost80}, it has been recognized that the QFIM and MUC are the real and imaginary parts of the so-called quantum Fisher tensor (QFT) \cite{Ercolessi12,Carollo18}, rendering a more unified picture for all these metrological quantities. 


In this paper, we advance this unified description of quantum metrology by proposing the generating functions for all the aforementioned metrological quantities. We elaborate that the logarithms of the Uhlmann relative amplitude between two density matrices\cite{Uhlmann86,Uhlmann91}, the modulus of the amplitude, and the phase of the amplitude are respectively the generating functions of QFT, QFIM, and MUC. These metrological quantities, as well as the Christoffel symbol of the first kind on the manifold of parameter space, are cumulants of the generating functions. Alternatively, the fidelity\cite{Nielsen10} between two density matrices can also serve as a generating function for the QFIM. In the historical development, this concept of generating function starts with the seminal work by Amari in probability theory\cite{Amari85}, which is then generalized to pure states described by quantum states $|\psi\rangle$ by Het\'{e}nyi and L\'{e}vay\cite{Hetenyi23,Hetenyi26}, and our theory further generalizes this concept to mixed states described by density matrices $\rho$. For pure states, our formalism recovers the logarithm of the product between two quantum states as the generating function for quantum geometric tensor\cite{Hetenyi23,Hetenyi26}, and clarifies the fidelity and Pancharatnam phase \cite{Pancharatnam56} of the product as the generating functions of quantum metric, Berry connection, and Berry curvature, which encode many quantum geometric and topological properties of a pure state, as demonstrated by 1D and 2D topological insulators.

Another advance of the present work concerns the Bloch representation of the density matrix, which is known to facilitate the practical calculation of the QFIM\cite{Watanabe10,Watanabe11,Liu20}. We formulate a Bloch representation for the QFT, which provides a generic expression for the tensor and MUC for any density matrix of arbitrary dimension, which dramatically simplifies the calculation of geometric phase for mixed states. Canonical ensembles of spin $1/2$ are used as an example to demonstrate the power of our technique, which helps to reveal many intriguing geometric properties of the three-dimensional (3D) manifold of magnetic field, including a constant Ricci scalar, vacuum Einstein equation, and a unity cosmological constant.

\section{Mixed state formalism}

\subsection{Generating functions for QFT \label{sec:gen_fn_mixed_state}}

Our discussion starts by considering a mixed state density matrix $\rho$ that depends on a set of $D$-dimensional parameters ${\bf x}=(x^{1},x^{2},\cdots,x^{D})$
\begin{align}
\rho({\bf x})=\sum_{i}\lambda_{i}({\bf x})|i({\bf x})\rangle\langle i({\bf x})|,
\end{align}
where we have expanded it in terms of the correctly normalized eigenstates $|i({\bf x})\rangle$ and eigenvalues $\lambda_{i}({\bf x})$ that satisfy $\sum_{i}\lambda_{i}({\bf x})=1$. The SLD $L_{\mu}$ is defined from 
\begin{align}
\partial_{\mu}\rho=\frac{1}{2}\left(\rho L_{\mu}+L_{\mu}\rho\right).
\label{SLD_definition}
\end{align}
where $\partial_{\mu}=\partial/\partial x^{\mu}$ denotes the derivative over a specific parameter. The metrological quantities under consideration are the QFT (or mixed state quantum geometric tensor) $T_{\mu\nu}$, the QFIM which is equal to four times the mixed state quantum metric $F_{\mu\nu}=4g_{\mu\nu}$\cite{Facchi10}, and MUC ${\cal U}_{\mu\nu}$ defined from the SLD by
\begin{align}
T_{\mu\nu} &=\frac{1}{4}{\rm Tr}\rho L_{\mu}L_{\nu}=\frac{1}{4}F_{\mu\nu}-\frac{i}{2}{\cal U}_{\mu\nu},
\nonumber \\
F_{\mu\nu} &=4g_{\mu\nu}=\frac{1}{2}{\rm Tr}\rho\left\{L_{\mu},L_{\nu}\right\},
\nonumber \\
{\cal U}_{\mu\nu} &=\frac{i}{4}{\rm Tr}\rho\left[L_{\mu},L_{\nu}\right],
\label{QFIM_matrix_element}
\end{align}
The prefactors of these quantities are defined in such a way that they descend to their pure state counterparts, as we shall see in Sec.~\ref{sec:pure_state_limit}. Note that an alternative mixed-state quantum geometric tensor defined from the density matrix and its purification has been proposed\cite{Hou24}, but we will use the $T_{\mu\nu}$ in Eq.~(\ref{QFIM_matrix_element}) to develop our theory.



Our aim is to construct a generating function for the QFT based on the theory of Uhlmann that expresses the density matrix in terms of the amplitudes\cite{Uhlmann86,Uhlmann91}, which we briefly review below. Consider the decomposition of a density matrix $\rho({\bf x})$ in terms of the amplitudes $w({\bf x})$
\begin{align}
\rho({\bf x})=w({\bf x})w^{\dag}({\bf x}).
\end{align}
The density matrix is invariant under the gauge transformation $w\rightarrow wU$. In practice, the amplitudes can be obtained from a gauge transformation over the square root of the density matrix $w=\sqrt{\rho}\,U$. Furthermore, two amplitudes $w$ and $w'$ of two density matrices $\rho$ and $\rho '$ are said to be parallel to each other, i.e., the parallel transport condition is fulfilled if $w^{\dag}w'=w^{'\dag}w\geq 0$\cite{Carollo18}. Thus if the second amplitude $w'$ is obtained from the first one $w$ via a small displacement $\delta x^{\mu}$ on the manifold $w'=w+\partial_{\mu}w\delta x^{\mu}$, then the parallel transport condition implies
\begin{align}
w^{\dag}\partial_{\mu}w=\partial_{\mu}w^{\dag}w.
\label{wdw_dww}
\end{align}
which is satisfied by the differential equations
\begin{align}
\partial_{\mu}w=\frac{1}{2}L_{\mu}w,\quad
\partial_{\mu}w^{\dag}=\frac{1}{2}w^{\dag}L_{\mu}.
\label{dw_Lw_dw_wL}
\end{align}
The question is then how to solve for the amplitude $w$ that satisfies this parallel transport condition for a given density matrix $\rho$.

To solve for the amplitude $w=\sqrt{\rho}\,U$ at a given $\sqrt{\rho}$, we begin by the ansatz for the derivative of $\sqrt{\rho}$\cite{Carollo18} 
\begin{align}
\partial_{\mu}\sqrt{\rho}=\frac{1}{2}L_{\mu}\sqrt{\rho}-i\sqrt{\rho}A_{\mu},
\label{dw_with_A}
\end{align}
which introduces the non-Abelian vector potential $A_{\mu}$. The problem now reduces to finding the appropriate local transformation $U$ from a given $A_{\mu}$ such that Eq.~(\ref{dw_Lw_dw_wL}) is satisfied. To solve for $U$, we take a derivative on $w=\sqrt{\rho}\,U$
\begin{align}
\partial_{\mu}w=\partial_{\mu}\sqrt{\rho}\,U+\sqrt{\rho}\,\partial_{\mu}U=\frac{1}{2}L_{\mu}w
-i\sqrt{\rho}A_{\mu}U+\sqrt{\rho}\,\partial_{\mu}U.
\end{align}
The last two terms must be cancel out for Eq.~(\ref{dw_Lw_dw_wL}) to be satisfied, leading to the differential equation for $U$
\begin{align}
\partial_{\mu}U=iA_{\mu}U,
\end{align}
whose solution is the Wilson line in non-Abelian gauge theories\cite{Peskin95}
\begin{align}
U_{\gamma}({\bf x})={\cal P}e^{i\int_{{\bf x}_{0},\gamma}^{{\bf x}}A_{\mu}({\bf y})\,dy^{\mu}}U({\bf x}_{0}),
\end{align}
where ${\cal P}$ denotes the path ordering, and ${\bf y}$ is the position on the trajectory $\gamma$ going from a reference point ${\bf x}_{0}$ to ${\bf x}$ (e.g., $y^{\mu}(s)=x_{0}^{\mu}+s(x^{\mu}-x_{0}^{\mu})$ with $0\leq s\leq 1$ if the trajectory $\gamma$ is a straight line). Consequently,
\begin{align}
w({\bf x})=\sqrt{\rho({\bf x})}\,{\cal P}e^{i\int_{{\bf x}_{0},\gamma}^{{\bf x}}A_{\mu}({\bf y})\,dy^{\mu}}U({\bf x}_{0}),
\label{w_parallel_transport_solution}
\end{align}
is the solution to $w$ that satisfies Eq.~(\ref{dw_Lw_dw_wL}).


We now demonstrate that under parallel transport condition, the trace of the Uhlmann relative amplitude $w({\bf x})w^{\dag}({\bf x'})\equiv ww'^{\dag}$ between the two density matrices $\rho({\bf x})\equiv\rho$ and $\rho({\bf x}')\equiv\rho '$, or logarithm of the trace, is a generating function for the QFT and the associated geometric properties. To formulate this, we first introduce the moments/cumulants defined from
\begin{align}
C_{\mu_{1}\mu_{2}...;\nu_{1}\nu_{2}...}\equiv (-i\partial_{\mu_{1}'})(-i\partial_{\mu_{2}'})...(i\partial_{\nu_{1}})(i\partial_{\nu_{2}})...
\left(\ln\right){\rm Tr}\,ww'^{\dag}|_{{\bf x=x'},},
\label{cumulant_Trww}
\end{align}
where the parenthesis $\left(\ln\right)$ means that one obtains the same quantity with or without taking a logarithm. The subscript ${\bf x=x'}$ in Eq.~(\ref{cumulant_Trww}) means that after taking the derivatives of $(-i\partial_{\mu_{1}'})(-i\partial_{\mu_{2}'})...$ on $w'^{\dag}$ and $(i\partial_{\nu_{1}})(i\partial_{\nu_{2}})...$ on $w$, one sets the two parameters to be the same ${\bf x=x'}$ and the two trajectories to be the same $\gamma=\gamma '$. Further using Eq.~(\ref{w_parallel_transport_solution}), the generating function takes the explicit form
\begin{align}
{\rm Tr}\,ww'^{\dag}={\rm Tr}\sqrt{\rho}\,{\cal P}e^{i\int_{{\bf x}_{0},\gamma}^{{\bf x}}A_{\mu}({\bf y})\,dy^{\mu}}{\cal P}e^{-i\int_{{\bf x}_{0},\gamma '}^{{\bf x'}}A_{\nu}({\bf y})\,dy^{\nu}}\sqrt{\rho '}.
\label{Trww_explicit_form}
\end{align}
Note that the factor $U^{\dag}({\bf x}_{0})U({\bf x}_{0})={\mathbb I}$ drops out, so the transformation at the starting point $U({\bf x}_{0})$ is unimportant in practice. This generating function ${\rm Tr}\,ww'^{\dag}$ is gauge-dependent in the sense that $w$ depends on the trajectories $\gamma$ that goes from ${\bf x}_{0}$ to ${\bf x}$ and $w'^{\dag}$ depends on the trajectory $\gamma '$ that goes from ${\bf x}_{0}$ to ${\bf x'}$. Nevertheless, the resulting cumulants in Eq.~(\ref{cumulant_Trww}) are gauge-invariant since we set ${\bf x=x'},\gamma=\gamma '$ after taking derivatives.

The first moments of Eq.~(\ref{cumulant_Trww}) vanish
\begin{align}
C_{\mu;}=C_{;\mu}=\frac{1}{2}{\rm Tr}\rho L_{\mu}=0.
\end{align}
It is because of these vanishing first moments that the moments obtained from taking derivatives on ${\rm Tr}\,ww'^{\dag}$ are the same as the cumulants obtained from taking derivatives on $\ln{\rm Tr}\,ww'^{\dag}$, since their difference consists of terms that are always multiplied by the first moment. Thus there is no difference between moments and cumulants, so we simply call $C_{\mu_{1}\mu_{2}...;\nu_{1}\nu_{2}...}$ the cumulants. Interestingly, all three second cumulants give the QFT (note the order of the indices)
\begin{align}
C_{\nu\mu;}=C_{\mu;\nu}=C_{;\mu\nu}=T_{\mu\nu},
\label{2nd_cumulant_Tmunu}
\end{align}
indicating the relative amplitude as a generating function for the QFT. Furthermore, from Eq.~(\ref{QFIM_matrix_element}) one sees that because QFIM and MUC are the real and imaginary parts of the QFT, their generating functions can be constructed from the modulus and phase of ${\rm Tr}\,ww'^{\dag}$ by
\begin{align}
&F_{\mu\nu} =(-i\partial_{\mu '})(i\partial_{\nu})\left[2{\rm Tr}\,ww'^{\dag}+2{\rm Tr}\,w'w^{\dag}\right]_{{\bf x=x'},}
\nonumber \\
&=(-i\partial_{\mu '})(i\partial_{\nu})\left[4{\rm Re}{\rm Tr}\,ww'^{\dag}\right]_{{\bf x=x'},}
\nonumber \\
&=(-i\partial_{\mu '})(i\partial_{\nu})\left[2\ln{\rm Tr}\,ww'^{\dag}+2\ln{\rm Tr}\,w'w^{\dag}\right]_{{\bf x=x'},}
\nonumber \\
&=(-i\partial_{\mu '})(i\partial_{\nu})\left[4\ln|{\rm Tr}\,ww'^{\dag}|\right]_{{\bf x=x'},}, 
\nonumber \\
&{\cal U}_{\mu\nu} = (-i\partial_{\mu '})(i\partial_{\nu})\left[i{\rm Tr}\,ww'^{\dag}-i{\rm Tr}\,w'w^{\dag}\right]_{{\bf x=x'},}
\nonumber \\
&=(-i\partial_{\mu '})(i\partial_{\nu})\left[(-2){\rm Im}{\rm Tr}\,ww'^{\dag}\right]_{{\bf x=x'},}
\nonumber \\
&=(-i\partial_{\mu '})(i\partial_{\nu})\left[i\ln{\rm Tr}\,ww'^{\dag}-i\ln{\rm Tr}\,w'w^{\dag}\right]_{{\bf x=x'},}
\nonumber \\
&=(-i\partial_{\mu '})(i\partial_{\nu})(-2)\varphi({\bf x,x'})|_{{\bf x=x'},}, 
\end{align}
with the phase $\varphi({\bf x,x'})$ introduced from ${\rm Tr}\,w({\bf x})w^{\dag}({\bf x'})=|{\rm Tr}\,w({\bf x})w^{\dag}({\bf x'})|e^{i\varphi({\bf x,x'})}$. 



The derivative on the QFT can also be expressed in terms of a combination of the third cumulants
\begin{align}
&&\partial_{\sigma}T_{\mu\nu}=i\left(C_{\sigma\mu;\nu}-C_{\mu;\sigma\nu}\right)
=\frac{1}{8}{\rm Tr}\rho \left\{L_{\sigma},L_{\mu}L_{\nu}\right\}+\frac{1}{4}{\rm Tr}\rho\partial_{\sigma}\left(L_{\mu}L_{\nu}\right).
\end{align}
As a result, if we treat the QFIM as a metric, then the corresponding Christoffel symbol of the first kind can be expressed in terms of the third cumulants
\begin{align}
&\Gamma_{\sigma\mu\nu} =\frac{1}{2}\left(\partial_{\nu}F_{\sigma\mu}
+\partial_{\mu}F_{\sigma\nu}-\partial_{\sigma}F_{\mu\nu}\right)=2{\rm Re}\left(\partial_{\nu}T_{\sigma\mu}
+\partial_{\mu}T_{\sigma\nu}-\partial_{\sigma}T_{\mu\nu}\right)
\nonumber \\
&=-2\,{\rm Im}\left(C_{\nu\sigma;\mu}-C_{\sigma;\nu\mu}+C_{\mu\sigma;\nu}-C_{\sigma;\mu\nu}
-C_{\sigma\mu;\nu}+C_{\mu;\sigma\nu}\right).
\end{align}
Finally, the second derivative of the QFT can be expressed in terms of the fourth cumulants
\begin{align}
&\partial_{\alpha}\partial_{\sigma}T_{\mu\nu} = C_{\sigma\mu;\alpha\nu} + C_{\alpha\mu;\sigma\nu} - C_{\alpha\sigma\mu;\nu} - C_{\mu;\alpha\sigma\nu} 
\nonumber \\
&=\frac{1}{16}\mathrm{Tr}[\rho\{L_{\alpha},\{L_{\sigma},L_{\mu}L_{\nu}\}\}] + \frac{1}{4}\mathrm{Tr}[\rho\partial_{\alpha}\partial_{\sigma}(L_{\mu}L_{\nu})] \nonumber\\
    &+ \frac{1}{8}\mathrm{Tr}[\rho\partial_{\alpha}\{L_{\sigma},L_{\mu}L_{\nu}\}]
    + \frac{1}{8}\mathrm{Tr}[\rho\{L_{\alpha},\partial_{\sigma}(L_{\mu}L_{\nu})\}].
\end{align}
As a result, the basic differential geometric properties such as the Christoffel symbol of the second kind $\Gamma_{\mu\nu}^{\lambda}$, Riemann tensor $R^{\rho}_{\;\sigma\mu\nu}$, Ricci tensor $R_{\mu\nu}$, Ricci scalar $R$, and Einstein tensor $G_{\mu\nu}$ defined by\cite{Carroll03}
\begin{align}
&\Gamma_{\mu\nu}^{\lambda} =F^{\lambda\sigma}\Gamma_{\sigma\mu\nu}=\frac{1}{2}F^{\lambda\sigma}(\partial_{\mu}F_{\nu\sigma} +\partial_{\nu}F_{\sigma\mu}-\partial_{\sigma}F_{\mu\nu}),
\nonumber \\
&R^{\rho}_{\;\sigma\mu\nu} =\partial_{\mu}\Gamma_{\nu\sigma}^{\rho}-\partial_{\nu}\Gamma_{\mu\sigma}^{\rho} +\Gamma_{\mu\lambda}^{\rho}\Gamma_{\nu\sigma}^{\lambda}-\Gamma_{\nu\lambda}^{\rho}\Gamma_{\mu\sigma}^{\lambda},
\nonumber \\
&R_{\mu\nu}=R^{\lambda}_{\;\mu\lambda\nu},\quad
R=F^{\mu\nu}R_{\nu\mu},\quad G_{\mu\nu}=R_{\mu\nu}-\frac{R}{2}F_{\mu\nu}.
\label{Christoffel_Riemann_Ricci}
\end{align} 
can all be expressed as polynomials of the second and third cumulants of the generating function, although we omit the lengthy expressions for simplicity. 



Finally, we remark that from the well-known expansion of the Uhlmann fidelity between neighboring density matrices\cite{Jozsa94,Liu20}
\begin{align}
&&{\rm Tr}\sqrt{\sqrt{\rho({\bf x})} \rho({\bf x+\delta x}) \sqrt{\rho({\bf x})}}=1-\frac{1}{8}F_{\mu\nu}\delta x^{\mu}\delta x^{\nu},
\end{align}
we see that either the fidelity or its logarithm serves as a generating function for the QFIM, in the sense that the QFIM can be obtained by taking derivatives
\begin{align}
&\frac{1}{4}F_{\mu\nu} = \left[(-i \partial_{\mu '})(i \partial_\nu){\rm Tr}\sqrt{\sqrt{\rho({\bf x})} \rho({\bf x'})\sqrt{\rho({\bf x})}}\right]_{\bf x=x'} 
\nonumber \\
&= \left[(-i \partial_{\mu '})(i \partial_\nu)\ln{\rm Tr}\sqrt{\sqrt{\rho({\bf x})} \rho({\bf x'})\sqrt{\rho({\bf x})}}\right]_{\bf x=x'}= g_{\mu\nu},
\label{real_gen_fn_Fmunu}
\end{align}
since the QFIM is the Hessian matrix and leading order expansion of the fidelity, and the parallel transport condition is not needed.  





\subsection{Bloch representation for QFT \label{sec:Bloch_rep_quantum_metrology}}

The Bloch representation for the QFIM in arbitrary dimension has been clarified, which greatly simplifies the practical calculation of the QFIM\cite{Watanabe10,Watanabe11,Liu20}. In this section, we demonstrate that the same method can be used to obtain the QFT and MUC. The formalism starts by expressing a $D$-dimensional density matrix $\rho({\bf x})$ in the generic form
\begin{equation}
\rho ({\bf x}) = \frac{1}{D} \left( \mathbb{I} + \sqrt{\frac{D(D-1)}{2}} \, \bm{r}({\bf x}) \cdot \bm{\kappa} \right),
\label{rho_Bloch_rep}
\end{equation}
where $\bm{r}({\bf x}) = (r_1, r_2,\cdots)^{\mathrm{T}}$ is the Bloch vector with the condition $|\bm{r}|^2 \le 1$ that describes the dependence on the parameters ${\bf x}$, and $\bm{\kappa}= (\kappa_{1},\kappa_{2},\cdots)$ is a $(D^2-1)$-dimensional vector of $SU(D)$ generators. Note that the Bloch vector $\bm{r}({\bf x})$ in higher dimensions is constrained by algebraic relations\cite{Kimura03}, but this does not affect our calculation below that applies to any parametrization of the Bloch vector. The generators are traceless $\mathrm{Tr}(\kappa_i) = 0$ and orthogonal $\mathrm{Tr}(\kappa_i \kappa_j)=2\delta_{ij}$, and also satisfy
\begin{align}
\left\{\kappa_{i},\kappa_{j}\right\} &=\frac{4}{D} \delta_{ij} \, \mathbb{I}
+ \sum_{m=1}^{d^2-1} \mu_{ijm} \kappa_m ,
\nonumber \\
[\kappa_i, \kappa_j] &= i \sum_{m=1}^{d^2-1} \epsilon_{ijm} \kappa_m,
\label{Bloch_eq_C2}
\end{align}
 where $\mu_{ijl} = \frac{1}{2} \, \mathrm{Tr}\,\left( \left\{\kappa_i, \kappa_j\right\} \kappa_l \right)$ and $\epsilon_{ijl} = -\frac{i}{2}\,\mathrm{Tr}\,\left( [\kappa_i,\kappa_j]\kappa_l \right)$ are the symmetric and antisymmetric structure constants (note that $\epsilon_{ijl}$ is not necessarily the Levi-Citiva symbol). Expressing the SLD in terms of the generators
\begin{equation}
L_{\mu} = z_{\mu} \mathbb{I} + {\bf y}_{\mu} \cdot {\boldsymbol\kappa},
\end{equation}
one can solve for the scalar $z_{\mu}$ and the vector ${\bf y}_{\mu}$. Introducing the real symmetric matrix G and skew-symmetric matrix M
\begin{align}
    \label{G_and_M}
    G_{ij} &= \frac{1}{2}\,\mathrm{Tr}\,\left(\rho\left\{\kappa_i,\kappa_j\right\}\right) = \frac{2}{D}\delta_{ij} + c \sum_m \mu_{ijm} r_m 
 \\
    M_{ij} &= \frac{1}{2}{\rm Tr}(\rho[\kappa_i,\kappa_j]) = ic\sum_m \epsilon_{ijm}r_m,   
\end{align}
where $c= \sqrt{(D-1)/(2D)} $, the solutions to $z_{\mu}$ and ${\bf y}_{\mu}$ are known to be\cite{Liu20}
\begin{align}
z_{\mu}  = -2c\,{\bf y}_{\mu}^{\mathrm{T}}{\bm r},\quad\quad {\bf y}_{\mu}
= \left( \frac{1}{2c} G - 2c\,{\bm r}\,{\bm r}^{\mathrm{T}} \right)^{-1} \partial_{\mu}{\bm r}.
\label{Bloch_eq_C17}
\end{align}
In this regard, one can calculate the QFT directly by observing that
\begin{align}
    {\rm Tr}\left(\rho L_\mu L_\nu\right) &= - z_\mu z_\nu  + \sum_{ij} y_{\mu,i}y_{\nu,j}{\rm Tr}(\rho \kappa_{i}\kappa_{j}),
\label{Trace_rhoLL}
\end{align}
where the last term in the foregoing equation can be simplified by Eq.~(\ref{G_and_M}) as
\begin{align}
   {\rm Tr}( \rho \kappa_i \kappa_j) = \frac{1}{2} {\rm Tr}( \rho\{\kappa_i,\kappa_j \}) + \frac{1}{2} {\rm Tr}( \rho[\kappa_i,\kappa_j]) = G_{ij} + M_{ij}.
\end{align}
Then, applying the above equation into Eq.~(\ref{Trace_rhoLL}), we obtain the expression for the QFT
\begin{align}
    T_{\mu\nu} = \frac{1}{4}{\rm Tr}\left(\rho L_\mu L_\nu\right) = -\frac{1}{4} z_\mu z_\nu + \frac{1}{4}\mathbf{y}_\mu^{T} (G + M) \mathbf{y}_\nu. 
    \label{QFT_Bloch_rep}
\end{align}
Furthermore, since the QFT can be decomposed into $T_{\mu\nu} = \frac{1}{4}F_{\mu\nu} - \frac{i}{2}{\cal U}_{\mu\nu}$, we see that the real part gives the well-known expression for the QFIM\cite{Liu20} 
\begin{align}
    F_{\mu\nu} &= 2c\,{\bf y}_{\mu}^{\mathrm{T}}\partial_{\nu}{\bm r}
    = (\partial_{\nu}{\bm r})^{\mathrm{T}} \left( \frac{D}{2(D-1)} G - {\bm r}\,{\bm r}^{\mathrm{T}} \right)^{-1} \partial_{\mu}{\bm r},
    \label{Bloch_eq_C21}
\end{align}
and the imaginary part gives the expression for the MUC
\begin{align}
    {\cal U}_{\mu\nu} &=  \frac{i}{2} \mathbf{y}_\mu^{T} M \mathbf{y}_\nu
    = \frac{i}{2}(\partial_\mu \bm{r})^T\ \left(\frac{1}{2c}G - 2c~\bm{r}\bm{r}^{T} \right)^{-1}  M\left(\frac{1}{2c}G - 2c~\bm{r}\bm{r}^{T} \right)^{-1}(\partial_\nu \bm{r}).
    \label{MUC_Bloch_rep}
\end{align}

Equations (\ref{QFT_Bloch_rep}) and (\ref{MUC_Bloch_rep}) are our main results for the Bloch representation, which completely express the QFT and MUC in terms of the derivatives on the Bloch vector ${\bm r}$ that characterizes the density matrix in Eq.~(\ref{rho_Bloch_rep}).




\subsection{Pure state limit \label{sec:pure_state_limit}}

For pure states ${\rm Tr}\rho^{2}=1$, the formalism dramatically simplifies because the gauge field $A_{\mu}$ becomes Abelian and can be made implicit. The density matrix in pure states is the same as its square root $\rho=\sqrt{\rho}=|\psi\rangle\langle\psi|$ that contains only a single eigenstate $|\psi\rangle$, so the amplitude $w$ is
\begin{align}
w=\sqrt{\rho}\,U=|\psi\rangle\langle\tilde{\psi}|,\quad
w^{\prime\dag}=U^{\prime\dag}\sqrt{\rho}=|\tilde{\psi}'\rangle\langle\psi '|,
\label{relative_amplitude_pure_state}
\end{align}
where $|\tilde{\psi}\rangle$ and $|\tilde{\psi}'\rangle$ are the states after the transformation $U=U({\bf x})$ and $U'=U({\bf x'})$. The SLD is given by\cite{Liu20} 
\begin{align}
L_{\mu}=2|\partial_{\mu}\psi\rangle\langle\psi|+2|\psi\rangle\langle\partial_{\mu}\psi|.
\end{align}
As a result, if we demand the parallel transport condition in Eq.~(\ref{dw_Lw_dw_wL}) to be satisfied, the transformed state $|\tilde{\psi}\rangle$ has to satisfy the differential equation
\begin{align}
|\partial_{\mu}\tilde{\psi}\rangle=|\tilde{\psi}\rangle\langle\psi|\partial_{\mu}\psi\rangle.
\label{dmutildepsi_differential}
\end{align}
From Eq.~(\ref{relative_amplitude_pure_state}), the relative amplitude reads
\begin{align}
{\rm Tr}\,ww^{\prime\dag}=\langle\psi '|\psi\rangle\langle\tilde{\psi}|\tilde{\psi}'\rangle,
\end{align}
whose second moment yields the quantum geometric tensor
\begin{align}
T_{\mu\nu} &= (-i\partial_{\mu '})(i\partial_{\nu}){\rm Tr}\,ww^{\prime\dag}|_{{\bf x=x'}}
=\langle\partial_{\mu}\psi|\partial_{\nu}\psi\rangle
-\langle\partial_{\mu}\psi|\psi\rangle\langle\psi|\partial_{\nu}\psi\rangle,
\end{align}
that correctly captures the pure state limit of the QFT. Interestingly, if we use the logarithm version of this equation, then
\begin{align}
&T_{\mu\nu} = (-i\partial_{\mu '})(i\partial_{\nu})\ln{\rm Tr}\,ww^{\prime\dag}|_{{\bf x=x'}}
=(-i\partial_{\mu '})(i\partial_{\nu})\left[\ln\langle\psi '|\psi\rangle+\ln\langle\tilde{\psi}|\tilde{\psi}'\rangle\right]_{{\bf x=x'}}
\nonumber \\
&=(-i\partial_{\mu '})(i\partial_{\nu})\left[\ln\langle\psi '|\psi\rangle\right]_{{\bf x=x'}}=\frac{\langle\partial_{\mu}\psi|\partial_{\nu}\psi\rangle}{\langle\psi|\psi\rangle}-
\frac{\langle\partial_{\mu}\psi|\psi\rangle\langle\psi|\partial_{\nu}\psi\rangle}{(\langle\psi|\psi\rangle)^{2}} ,
\end{align}
which recovers the $\ln\langle\psi '|\psi\rangle$ as the generating function for the quantum geometric tensor proposed by Het\'{e}nyi and L\'{e}vay\cite{Hetenyi23}, where we have used $(-i\partial_{\mu '})(i\partial_{\nu})[\ln\langle\tilde{\psi}|\tilde{\psi}'\rangle]_{{\bf x=x'}}=0$ according to Eq.~(\ref{dmutildepsi_differential}). Thus the generating function $\ln\langle\psi '|\psi\rangle$ proposed by Het\'{e}nyi and L\'{e}vay has a profound meaning as the logarithm of the relative amplitude between two density matrices, and is the pure state limit of the more general generating function $\ln{\rm Tr}\,ww'^{\dag}$ proposed in the present work. Finally, we remark that the generating function $\ln\langle\psi '|\psi\rangle$ proposed by Het\'{e}nyi and L\'{e}vay is also gauge-dependent: Under the local $U(1)$ transformation $|\psi({\bf x})\rangle\rightarrow e^{-i\varphi({\bf x})}|\psi({\bf x})\rangle$ that leaves the density matrix $\rho({\bf x})$ unchanged, the generating function transforms like $\ln\langle\psi({\bf x'})|\psi({\bf x})\rangle\rightarrow\ln\langle\psi({\bf x'})|\psi({\bf x})\rangle+i\varphi({\bf x'})-i\varphi({\bf x})$. Nevertheless, the resulting cumulants are gauge-invariant, just like in the mixed-state cases discussed after Eq.~(\ref{Trww_explicit_form}).

Alternatively, the fidelity itself $|\langle\psi({\bf x'})|\psi({\bf x})\rangle|\equiv|\langle\psi '|\psi\rangle|$ or its logarithm can also serve as a generating function for the quantum metric, since 
\begin{align}
g_{\mu\nu} &= (-i \partial_{\mu '})(i \partial_{\nu})\sqrt{\langle\psi '|\psi\rangle\langle\psi|\psi '\rangle}|_{\bf x=x'} =\frac{1}{2}\partial_{\mu '}\left[\frac{\langle\psi '|\partial_{\nu}\psi\rangle\langle\psi|\psi '\rangle+\langle\psi '|\psi\rangle\langle\partial_{\nu}\psi|\psi '\rangle}{\sqrt{\langle\psi '|\psi\rangle\langle\psi|\psi '\rangle}}\right]_{\bf x=x'},
\label{gen_fn_gmunu_without_log}
\end{align}
which is simply the pure state limit of Eq.~(\ref{real_gen_fn_Fmunu}) that needs not invoke parallel transport. Besides, one also has the following obvious generating function for the Berry curvature 
\begin{align}
&\Omega_{\mu\nu} = i(-i \partial_{\mu '})(i \partial_\nu)\left[\langle\psi({\bf x'})|\psi({\bf x})\rangle-\langle\psi({\bf x})|\psi({\bf x'})\rangle\right]_{\bf x=x'}
\nonumber \\
&=(-i \partial_{\mu '})(i \partial_\nu)\left[-2{\rm Im}\langle\psi({\bf x'})|\psi({\bf x})\rangle\right]_{\bf x=x'}.
\label{gen_fn_Omegamunu_without_log}
\end{align}
Combing Eqs.~(\ref{gen_fn_gmunu_without_log}) and (\ref{gen_fn_Omegamunu_without_log}) yields an alternative generating function for $T_{\mu\nu}$
\begin{align}
T_{\mu\nu}=(-i \partial_{\mu '})(i \partial_\nu)\left[|\langle\psi({\bf x'})|\psi({\bf x})\rangle|+i{\rm Im}\langle\psi({\bf x'})|\psi({\bf x})\rangle\right]_{\bf x=x'},
\end{align}
that works equally well as the $\ln\langle\psi '|\psi\rangle$ proposed by Het\'{e}nyi and L\'{e}vay\cite{Hetenyi23}. Finally, differentiation on the Pancharatnam phase $\varphi({\bf x',x})$ of the product $\langle\psi({\bf x'})|\psi({\bf x})\rangle=|\langle\psi({\bf x'})|\psi({\bf x})\rangle|\,e^{i\varphi({\bf x',x})}$ gives
\begin{align}
&\Omega_{\mu\nu}= i(-i \partial_{\mu '})(i \partial_\nu)\left[\ln\langle\psi({\bf x'})|\psi({\bf x})\rangle-\ln\langle\psi({\bf x})|\psi({\bf x'})\rangle\right]_{\bf x=x'}
\nonumber \\
&=i\partial_{\mu '}\partial_{\nu}\ln e^{2i\varphi({\bf x',x})}|_{\bf x=x'}=\partial_{\mu '}\partial_{\nu}\left[-2\varphi({\bf x',x})\right]_{\bf x=x'},
\nonumber \\
&2i{\cal A}_{\mu} =-i\partial_{\mu '}\left[-2\varphi({\bf x',x})\right]_{\bf x=x'},
\nonumber \\
&2i{\cal A}_{\nu} =i\partial_{\nu}\left[-2\varphi({\bf x',x})\right]_{\bf x=x'},
\label{gen_fn_Omegamunu_phase}
\end{align}
so the Pancharatnam phase $\varphi({\bf x',x})$ also serves as a generating function, whose first cumulant gives the Berry connection ${\cal A}_{\mu}=\langle\psi|i\partial_{\mu}|\psi\rangle$ and second cumulant the Berry curvature $\Omega_{\mu\nu}$.

The second, third, and fourth cumulants obtained from the generating function in the pure state limit are still those given in Sec.~\ref{sec:gen_fn_mixed_state}. However, in pure states, it is customary to use the quantum metric $g_{\mu\nu}$ instead of the QFIM $F_{\mu\nu}=4g_{\mu\nu}$ as the metric to define differential geometric properties, thus a factor of four different from those in Sec.~\ref{sec:gen_fn_mixed_state}. With this factor of four included, they are given by
\begin{align}
&\Gamma_{\lambda\mu\nu}=\frac{1}{2}\left[\partial_{\nu}g_{\lambda\mu}+\partial_{\mu}g_{\lambda\nu}
-\partial_{\lambda}g_{\mu\nu}\right]
\nonumber \\
&=-\frac{1}{2}\,{\rm Im}\left(C_{\nu\sigma;\mu}-C_{\sigma;\nu\mu}+C_{\mu\sigma;\nu}-C_{\mu;\sigma\nu}
-C_{\sigma\mu;\nu}+C_{\mu;\sigma\nu}\right)
\nonumber \\
&=\frac{1}{2}\left\{
\langle \partial_\lambda \psi | (I - |\psi\rangle\langle\psi|) | \partial_\mu \partial_\nu \psi \rangle
+ \langle \partial_\mu \partial_\nu \psi | (I - |\psi\rangle\langle\psi|) | \partial_\lambda \psi \rangle \right.
\nonumber \\
& + \left.\langle \partial_\mu \psi | (|\partial_\lambda \psi\rangle\langle\psi| + |\psi\rangle\langle \partial_\lambda \psi|) | \partial_\nu \psi \rangle 
+ \langle \partial_\nu \psi | (|\partial_\lambda \psi\rangle\langle\psi| + |\psi\rangle\langle \partial_\lambda \psi|) | \partial_\mu \psi \rangle \right\},
\nonumber \\
&\partial_{\alpha}\partial_{\sigma}g_{\mu\nu} ={\rm Re}\,\partial_{\alpha}\partial_{\sigma}T_{\mu\nu} = {\rm Re}\left(C_{\sigma\mu;\alpha\nu} + C_{\alpha\mu;\sigma\nu} - C_{\alpha\sigma\mu;\nu} - C_{\mu;\alpha\sigma\nu} \right).
\label{Christoffel_symbol_cumulant_expression}
\end{align}
where the expression for $\Gamma_{\lambda\mu\nu}$ in terms of $|\psi\rangle$ has been derived recently\cite{Pal26}.



\subsection{\texorpdfstring{$2\times2$}{2x2} density matrices \label{sec:2by2_density_mat}}

The $2\times 2$ density matrices are fundamental to describe many thermodynamic properties of countless two-state quantum systems, whose Bloch representation certainly deserves to be formulated in detail. With the parametrization $\rho=({\mathbb I}_{2}+{\bm r}\cdot{\boldsymbol\sigma})/2$ and $D=2$, $\kappa_i$ are the Pauli matrices  $\kappa_{i}=\sigma_{i}$, satisfying $\left\{\sigma_{i},\sigma_{j}\right\}=2\delta_{ij}{\mathbb I}_{2}$, and $c=1/2$, so $G={\mathbb I}_{3}$. The structure constant $\epsilon_{ijl}$ becomes two times the Levi-Civita symbol. Using the formulas in Sec.~\ref{sec:Bloch_rep_quantum_metrology}, one obtains the well-known result for the QFIM\cite{Liu20}, and an expression for the MUC
\begin{align}
F_{\mu\nu} &=\partial_{\mu}{\bm r}\cdot\partial_{\nu}{\bm r}
+\frac{({\bm r}\cdot\partial_{\mu}{\bm r})({\bm r}\cdot\partial_{\nu}{\bm r})}{1-|{\bm r}|^{2}},
\nonumber \\
{\cal U}_{\mu\nu} &=-\frac{1}{2}{\bm r}\cdot\partial_{\mu}{\bm r}\times\partial_{\nu}{\bm r}.
\label{QFIM_MUC_2by2_Bloch}
\end{align}
We shall see some applications of these formulas in Sec.~\ref{sec:spin_half}. The Uhlmann fidelity in this case is known to be\cite{Jozsa94} 
\begin{align}
&&{\rm Tr}\sqrt{\sqrt{\rho({\bf x})} \rho({\bf x'})\sqrt{\rho({\bf x})}}
=\left[\frac{1}{2}(1+{\bm r}\cdot {\bm r'})+\frac{1}{2}\sqrt{1-|{\bm r}|^{2}}\sqrt{1-|{\bm r}^{\prime}|^{2}}\right]^{1/2}.
\label{fidelity_2by2_Bloch_rho}
\end{align}
Applying this fidelity to Eq.~(\ref{real_gen_fn_Fmunu}) indeed obtains the QFIM in Eq.~(\ref{QFIM_MUC_2by2_Bloch}), confirming the fidelity as a generating function of the QFIM.

The gauge field $A_{\mu}$ in Eq.~(\ref{dw_with_A}) can be solved by observing that $\sqrt{\rho}$ takes the form
\begin{align}
\sqrt{\rho}=a_+{\mathbb I}_{2}+a_-\frac{{\bm r}}{|{\bm r}|}\cdot{\boldsymbol\sigma},\qquad
a_\pm=\frac{\sqrt{\lambda_{+}} \pm\sqrt{\lambda_{-}}}{2},
\end{align}
where $\lambda_{\pm}=(1\pm|{\bm r}|)/2$ are the eigenvalues of the density matrix. By choosing the $SU(2)$ gauge $A_{\mu}={\bf A}_{\mu}\cdot{\boldsymbol\sigma}$, Eq.~(\ref{dw_with_A}) gives the solution 
\begin{align}
A_{\mu}=-\frac{({\bm r\times\partial_{\mu}{\bm r}})\cdot{\boldsymbol\sigma}}{2\left(1+\sqrt{1-|{\bm r}|^{2}}\right)},
\end{align}
and as a result
\begin{align}
{\rm Tr}\,ww'^{\dag} &={\rm Tr}\left[a_+{\mathbb I}_{2} + a_-\frac{{\bm r}}{|{\bm r}|}\cdot{\boldsymbol\sigma}\right]{\cal P}\exp\left\{-i\int_{\bf x_{0},\gamma}^{\bf x}\frac{({\bm r\times\partial_{\mu}{\bm r}})\cdot{\boldsymbol\sigma}}{2\left(1+\sqrt{1-|{\bm r}|^{2}}\right)}dy^{\mu}\right\}
\nonumber \\
&\quad\times{\cal P}\exp\left\{i\int_{\bf x_{0},\gamma '}^{\bf x'}\frac{({\bm r'\times\partial_{\nu}{\bm r'}})\cdot{\boldsymbol\sigma}}{2\left(1+\sqrt{1-|{\bm r'}|^{2}}\right)}dy^{\nu}\right\}\left[a_+'{\mathbb I}_{2}+a_-'\frac{{\bm r'}}{|{\bm r'}|}\cdot{\boldsymbol\sigma}\right],
\label{Trww_explicit_form_2by2}
\end{align}
is the Bloch representation for the generating function in Eq.~(\ref{Trww_explicit_form}).

For pure states, let us consider a generic $2\times 2$ Hamiltonian expanded by Pauli matrices $H({\bf x})={\bm d({\bf x})}\cdot{\boldsymbol\sigma}$, where ${\bm d({\bf x})}=(d_{1},d_{2},d_{3})$ describes the parameter-dependence and gives the eigenenergies $\pm d=\pm(d_{1}^{2}+d_{2}^{2}+d_{3}^{2})^{1/2}$. Introducing the unit vector ${\bm n}={\bm d}/d$, the density matrix of the eigenstate $|\psi\rangle$ with negative eigenenergy $-d$ is $\rho=|\psi\rangle\langle\psi|=\left({\mathbb I}_{2}-{\bm n}\cdot{\boldsymbol\sigma}\right)/2$. Denoting the $\left\{d_{i},n_{i}\right\}$ and $\left\{d_{i}',n_{i}'\right\}$ to be the functions at ${\bf x}$ and ${\bf x'}$, the generating functions and geometric quantities derived in Sec.~\ref{sec:pure_state_limit} are
\begin{align}
\label{Trpsi_gmunu_Gamma}
&(\ln)|\langle\psi({\bf x'})|\psi({\bf x})\rangle|=(\ln)\left[\frac{1}{2}(1+{\bm n'\cdot n})\right]^{1/2}=(\ln)\left|\frac{{\bm n'}+{\bm n}}{2}\right|,
\nonumber \\
&-2\varphi({\bf x',x})=-2\arg\left[\left(1-n_{3}-n_{3}'+{\bm n'}\cdot {\bm n}\right)+i\left(n_{1}'n_{2}-n_{2}'n_{1}\right)\right],
\nonumber \\
&-2{\rm Im}\langle\psi({\bf x'})|\psi({\bf x})\rangle
=\frac{n_{2'}n_{1}-n_{1}'n_{2}}{\sqrt{1-n_{3}'}\sqrt{1-n_{3}}},
\nonumber \\
&g_{\mu\nu}=\frac{1}{4}F_{\mu\nu}=\frac{1}{4}\partial_{\mu}{\bm n}\cdot\partial_{\nu}{\bm n}=\frac{1}{4d^{2}}\left(\sum_{i=1}^{2}\partial_{\mu}d_{i}\partial_{\nu}d_{i}-\partial_{\mu}d\partial_{\nu}d\right),
\nonumber \\
&\Omega_{\mu\nu}=\frac{1}{2}{{\bm n}\cdot\left(\partial_{\mu}{\bm n}\times\partial_{\nu}{\bm n}\right)}
=\frac{1}{2d^{3}}{{\bm d}\cdot\left(\partial_{\mu}{\bm d}\times\partial_{\nu}{\bm d}\right)},
\nonumber \\
&\Gamma_{\lambda\mu\nu}=\frac{1}{4}\partial_{\lambda}{\bm n}\cdot\partial_{\mu}\partial_{\nu}{\bm n},
\end{align}
recovering the well-known expressions for the quantum metric and Berry curvature for $2\times 2$ Hamiltonians\cite{Matsuura10,vonGersdorff21_metric_curvature,Xiao10}. It is also evident that Eqs.~(\ref{QFIM_MUC_2by2_Bloch}) and (\ref{fidelity_2by2_Bloch_rho}) recover Eq.~(\ref{Trpsi_gmunu_Gamma}) in the pure state limit ${\bm r}=-{\bm n}$ and ${\bm r}\cdot\partial_{\mu}{\bm r}={\bm n}\cdot\partial_{\mu}{\bm n}=0$. 

\section{Applications \label{sec:applications}}

\subsection{Canonical ensemble of spin-\texorpdfstring{$1/2$}{1/2} in a magnetic field \label{sec:spin_half}}

\begin{figure}[ht]
\centering
\includegraphics[width=0.9\textwidth]{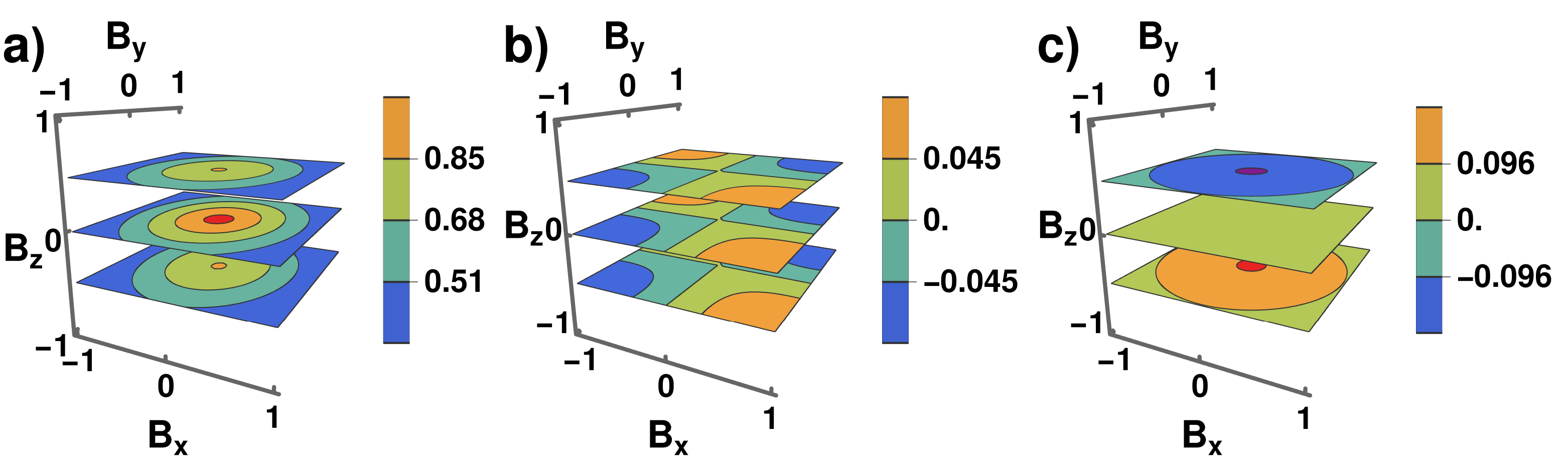}
\caption{Components of QFIM and MUC on the 3D magnetic field manifold for a single spin-$1/2$, where we plot (a) $F_{xx}$, (b) $F_{xy}$ and (c) $\mathcal{U}_{xy}$ at $\beta=1/k_{B}T=1$. Other components can be obtained by rotating these patterns in a way that respects the cubic symmetry. }
\label{fig:QFIM_and_MUC_spin-half}
\end{figure}

We proceed to use the canonical ensemble of a single spin-$1/2$ in a magnetic field to demonstrate the Bloch representation. Consider the $2\times 2$ Hamiltonian $H=-{\bm B}\cdot{\boldsymbol\sigma}$ with the magnetic field  ${\bm B}=(B_{x},B_{y},B_{z})$ pointing in an arbitrary direction, and $\beta=1/k_{B}T$ is determined by the temperature $T$. The density matrix is
\begin{align}
\rho=\frac{e^{-\beta H}}{{\rm Tr}e^{-\beta H}}=\frac{1}{2}\left(\mathbb{I}_{2}+{\bm r}\cdot{\boldsymbol\sigma}\right), \quad\quad{\bm r}=\frac{\bm B}{B}\tanh\beta B
\end{align}
where we denote $B=\sqrt{B_{x}^{2}+B_{y}^{2}+B_{z}^{2}}$ as the magnitude of the magnetic field. The three components of the magnetic field $(B_{x},B_{y},B_{z})$ are treated as a $3D$ Euclidean manifold with QFIM as the metric, so the differentiation is with respect to the magnetic field $\partial_{\mu}\equiv\partial/\partial B_{\mu}$. Using the Bloch representation in Sec.~\ref{sec:2by2_density_mat}, the generating function is exactly that given in Eq.~(\ref{Trww_explicit_form_2by2}), and the QFIM and MUC in Eq.~(\ref{QFIM_MUC_2by2_Bloch}) are found to be
\begin{align}
&F_{\mu\nu} =  \frac{\tanh^2(\beta B)}{B^2}\left( \delta_{\mu\nu} -\frac{B_\mu B_\nu}{B^2}\right) +   \beta^2 \text{sech}^2(\beta B) \frac{B_\mu B_\nu}{B^2}
\nonumber \\
&\det F_{\mu\nu} =\frac{\beta^{2}\,{\rm sech}^{2}\beta B\times \tanh^{4}\beta B}{B^{4}},
\nonumber \\
&{\cal U}_{\mu\nu} =-\frac{\varepsilon_{\mu\nu\sigma}B_{\sigma}\tanh^{3}\beta B}{2B^{3}},
\label{QFIM_MUC_spin_half}
\end{align}
where $\varepsilon_{\mu\nu\sigma}$ is the Levi-Civita symbol. This yields a constant total volume of the manifold
\begin{align}
V=\int d^{3}{\bm B}\sqrt{\det F_{\mu\nu}}=\pi^{2}.
\end{align}
Numerical results of the QFIM and MUC shown in Fig.~\ref{fig:QFIM_and_MUC_spin-half} indicate an eliptical shape of $F_{\mu\mu}$, and an $F_{\mu\neq\nu}$ and ${\cal U}_{\mu\nu}$ that change sign on the 3D manifold of magnetic field, reflecting the magnetic field dependence described by Eq.~(\ref{QFIM_MUC_spin_half}).

Inserting Eq.~(\ref{QFIM_MUC_spin_half}) into Eq.~(\ref{Christoffel_Riemann_Ricci}) reveals several remarkable geometric properties of the 3D magnetic field manifold. Firstly, the Ricci scalar is found to be a constant $R=6$, meaning that the whole manifold is curved in a uniform manner. Secondly, the Ricci tensor is twice the QFIM $R_{\mu\nu}=2F_{\mu\nu}$, rendering the Einstein tensor equal to the negative of the QFIM $G_{\mu\nu}=-F_{\mu\nu}$, and hence the vacuum Einstein equation 
\begin{align}
G_{\mu\nu}+\Lambda F_{\mu\nu}=0,
\end{align}
is satisfied by a unity cosmological constant $\Lambda=1$. These geometric properties are very similar to those discovered in the momentum space of 3D topological materials\cite{Chen25_generic_TI_TSC}, pointing to intriguing geometric properties associated with the QFIM and quantum metric in thermodynamic ensembles. 


\subsection{Su-Schrieffer-Heeger model}

We proceed to use the Su-Schrieffer-Heeger (SSH) model at zero temperature to elaborate the pure-state result in Sec.~\ref{sec:2by2_density_mat}. The model in real space is described by\cite{Su79}
\begin{align}
{\cal H}_{0} = \sum_{i}(t+\delta t)c_{Ai}^{\dag}c_{Bi}+(t-\delta t)c_{Ai+1}^{\dag}c_{Bi}+h.c.
\end{align}
where $c_{Ai}$ and $c_{Bi}$ are the fermion annihilation operators on sublattice $A$ and $B$ in the unit cell $i$, respectively, with alternating hopping amplitude $t\pm\delta t$ on even and odd bonds. In momentum space, the model is a $2\times 2$ Dirac model with
\begin{align}
    d_{1}=(t+\delta t)+(t-\delta t)\cos k,\quad d_{2}=(t-\delta t)\sin k,\quad d_{3}=0.
\end{align}
The model is known to undergo a topological phase transition at $\delta t_{c}=0$ accompanied by a divergence of quantum metric $g_{kk}$ and Berry connection ${\cal A}_{k}$ of the negative-energy state at $k=\pi$. To delineate the critical behavior, we choose to examine the parameters in the topologically nontrivial phase far from $\delta t=-0.5$ and close to $\delta=-0.1$ the critical point, and in the topologically trivial phase far from $\delta t=0.5$ and close to $\delta=0.1$ the critical point.

\begin{figure}[ht]
\centering
\includegraphics[width=0.99\textwidth]{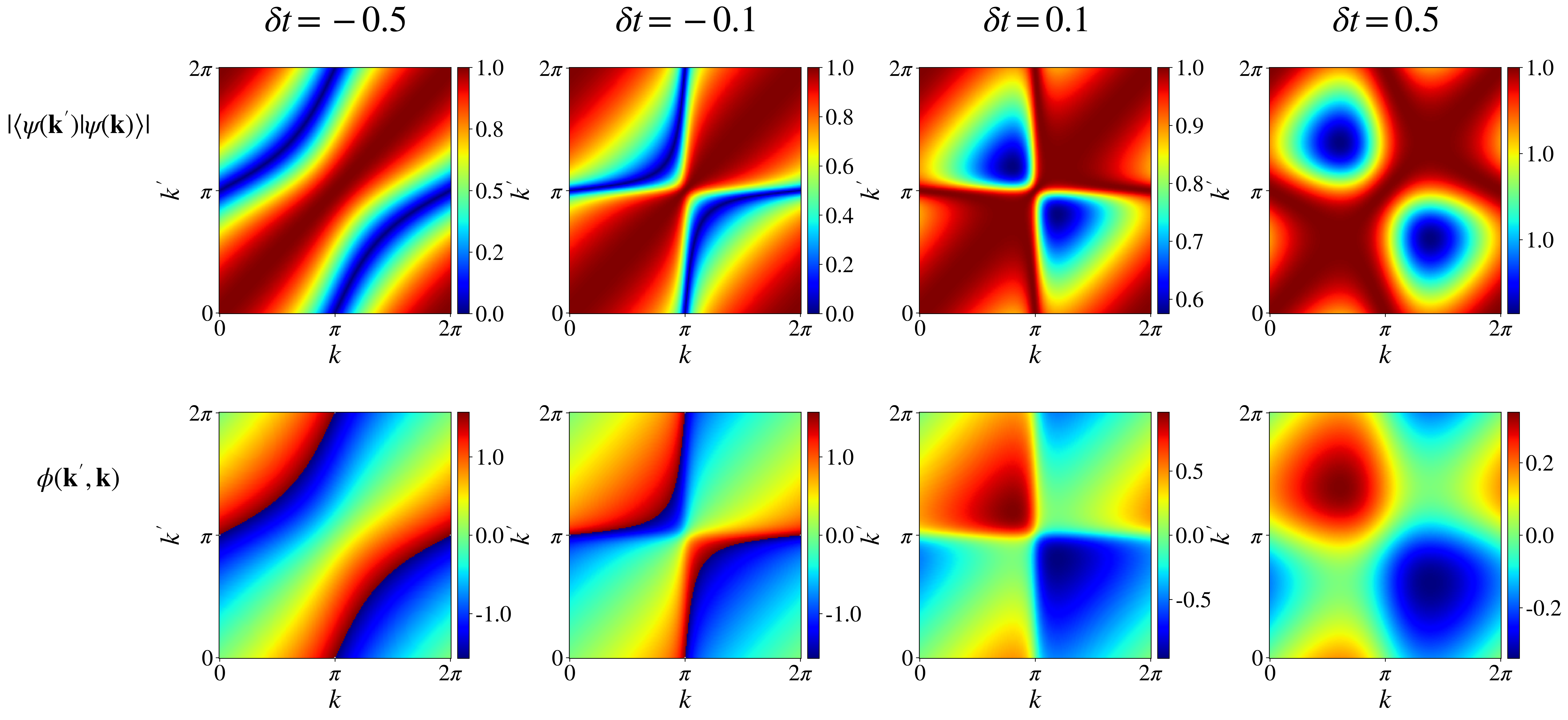}
\caption{Fidelity (top row) and Pancharatnam phase (bottom row) as functions of the two momenta $k$ and $k'$, in the topologically nontrivial phase far from $\delta t=-0.5$ and close to $\delta=-0.1$ the critical point, and in the topologically trivial phase far from $\delta t=0.5$ and close to $\delta=0.1$ the critical point.}
\label{fig:260929_Density_plot_SSH}
\end{figure}

The fidelity $|\langle\psi(k')|\psi(k)\rangle|$ and Pancharatnam phase $\varphi(k',k)$ as generating functions are presented as functions of the two momenta in Fig.~\ref{fig:260929_Density_plot_SSH}. One can visualize the quantum metric $g_{kk}$ as the second cumulant by taking the slope of the slope of the fidelity in the two directions $\partial_{k}\partial_{k'}|\langle\psi(k')|\psi(k)\rangle|$ evaluted along the diagonal $k=k'$ (higher cumulants can also be visualized by the same procedure). Similarly, Berry connection ${\cal A}_{k}$ as the first cumulant is the slope of the Pancharatnam phase in either direction $\partial_{k}\varphi(k',k)$ or $\partial_{k'}\varphi(k',k)$ evaluted along the diagonal $k=k'$. As a result, the divegence of quantum metric and Berry curvature are embedded in the fidelity and Pancharatnam phase, which show lobes of maxima close to $k=k'=\pi$ that gradually diverge as the system approaches the critical point $\delta t\rightarrow 0$, leading us to conclude that the generating functions encode all the quantum geometric and topological properties in this model.

\subsection{Chern insulator}

We proceed to use the Chern insulator at zero temperature as an example for 2D systems. The momentum space of Chern insulator is described by the spinless basis $\left(c_{{\bf k}s},\;c_{{\bf k}p}\right)^{T}$ and Hamiltonian\cite{Qi06,Qi11,Bernevig13}
\begin{align}
H({\bf k}) &= A\sin k_{x}\sigma^{x}+A\sin k_{y}\sigma^{y} +\left(M+4B-2B\cos k_{x}-2B\cos k_{y}\right)\sigma^{z},
\end{align} 
where we choose $A=B=1$ as energy unit. The negative-energy state is used to calculate the quantum metric and Berry curvature, with momentum ${\bf x\rightarrow k}$ treated as a 2D parameter space. Because the generating functions depend on the two momenta ${\bf k}=(k_{x},k_{y})$ and ${\bf k}'=(k_{x}',k_{y}')$, they are functions of four variables $(k_{x},k_{y},k_{x}',k_{y}')$ that is difficult to present. For this reason, we choose to plot the generating functions in the 2D subspace $(k_{x}=k_{y}',k_{y}=k_{x}')$ for the sake of presentation.

\begin{figure}[ht]
\centering
\includegraphics[width=0.99\textwidth]{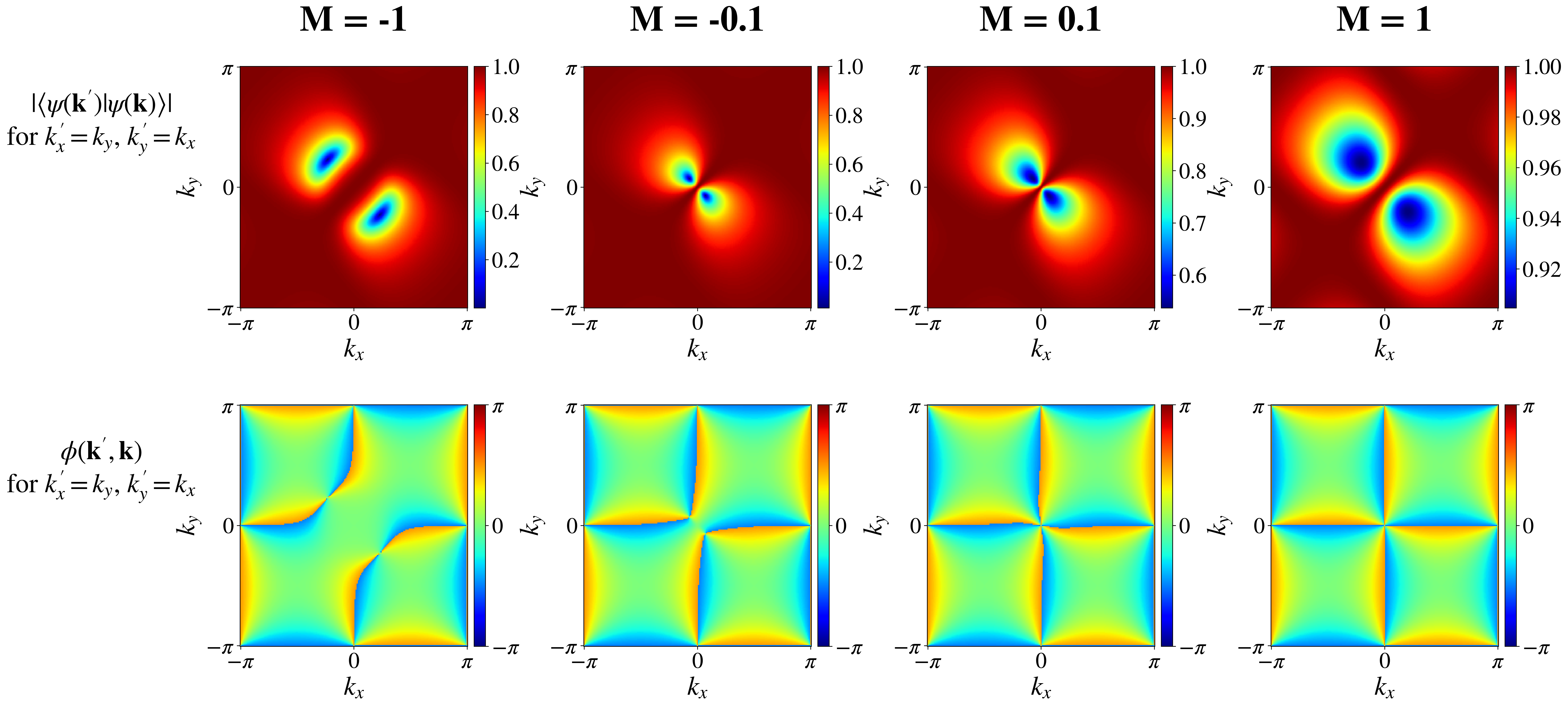}
\caption{Fidelity (top row) and Pancharatnam phase (bottom row) of the Chern insulator in the topologically nontrivial phase far from ($M=-1$) and close to ($M=-0.1$) the critical point, and in the topologically trivial phase far from ($M=1$) and close to ($M=0.1$) the critical point. These quantities are plotted in the 2D subspace $(k_{x}=k_{y}',k_{y}=k_{x}')$ of the 4D parameter space $(k_{x},k_{y},k_{x}',k_{y}')$ for the sake of demonstrating their features.}
\label{fig:260915_Density_plot_All}
\end{figure}

In Fig.~\ref{fig:260915_Density_plot_All}, we present the fidelity $|\langle\psi({\bf k'})|\psi({\bf k})\rangle|$ and Pancharatnam phase $\varphi({\bf k',k})$ in this 2D subspace in the topologically nontrivial phase far from ($M=-1$) and close to ($M=-0.1$) the critical point $M_{c}=0$ of a topological phase transition, as well as in the topologically trivial phase far from ($M=1$) and close to ($M=0.1$) the critical point, which manifest the following features. Firstly, at the critical point $M_{c}=0$, both the quantum metric $g_{\mu\nu}$ and quantum metric $\Omega_{xy}$ are known to diverge at momentum ${\bf k}=(0,0)$. The momentum profile of the fidelity exhibits two lobes around the center of the plot $k_{x}=k_{y}=k_{x}'=k_{y}'=0$, and the lobes gradually diverge as the system approaches the critical point (see $M=0.1$ and $M=-0.1$), in accordance with the divergence of quantum metric at ${\bf k}=(0,0)$ as the second cumulant of fidelity. The Pancharatnam phase also exhibits a similar divergence feature near $k_{x}=k_{y}=k_{x}'=k_{y}'=0$ (although the pattern in the topologically trivial phase is less sensitive to $M$ in this 2D subspace), indicating the divergence of Berry curvature at ${\bf k}=(0,0)$ as its second cumulant. These features again suggest that the fidelity and Pancharatnam phase as generating functions encode all the information about the quantum geometric and topological properties of the Chern insulator.


\section{Conclusions}

In summary, we elaborate the role of the Uhlmann relative amplitude between two mixed state density matrices as a generating function for quantum metrological properties. The QFT, QFIM, and MUC can be obtained through successive derivatives over system parameters on the logarithm, modulus, and phase of the relative amplitude, respectively. The Christoffel symbol and other differential geometric properties can also be obtained from higher cumulants of the generating function. Alternatively, the fidelity between two density matrices can also serve as a generating function for the QFIM. For pure states, our formalism recovers the generating function for quantum geometric tensor known in the literature, and also clarifies the fidelity between two quantum states as a generating function for the quantum metric, and the Pancharatnam phase between two quantum states as a generating function for the Berry connection and Berry curvature, offering a unified picture that encodes many quantum geometric and topological properties of condensed matter systems.

We further formulate a generic expression for the QFT and MUC based on the Bloch representation of density matrices of any dimension, which facilitates the calculation of these quantities in practice. The investigation on spin-$1/2$ reveals a constant Ricci scalar, vacuum Einstein equation, and a unity cosmological constant on the 3D manifold of magnetic field, suggesting the Bloch representation as a powerful tool to address the differential geometric properties and geometric phase of mixed states.

\section*{Acknowledgments}
The authors acknowledge the support of the INCT project Advanced Quantum Materials, involving the Brazilian agencies CNPq (Proc. 408766/2024-7), FAPESP, and CAPES, as well as the financial support of the productivity in research fellowship from CNPq. 

\appendix 


\section*{References}
\bibliographystyle{iopart-num}
\bibliography{Literatur}

\end{document}